# Depth-resolved holographic reconstructions by three-dimensional volumetric deconvolution

Tatiana Latychevskaia[*], Fabian Gehri and Hans-Werner Fink

*Physics Institute, University of Zurich, Winterthurerstrasse 190, 8057 Zurich, Switzerland*
*Corresponding author: tatiana@physik.uzh.ch

## Abstract

Methods of three-dimensional volumetric deconvolution with a point-spread function as frequently employed in optical microscopy to reconstruct true three-dimensional distribution of objects are extended to holographic reconstructions. Two such schemes have been developed and are discussed: an instant deconvolution as well as an iterative deconvolution routine. The instant three-dimensional volumetric deconvolution can be applied to restore the positions of volume-spread objects such as small particles. The iterative deconvolution can be applied to restore the distribution of complex and extended objects. Simulated and experimental examples are presented and demonstrate artifact and noise free three-dimensional reconstructions from a single two-dimensional holographic record.

The method of three-dimensional volumetric deconvolution is applied to the reconstructed wavefronts. The method is universal as it does not depend on how the reconstructed wavefronts were obtained, by employing Huygens-Fresnel diffraction, Rayleigh-Sommerfeld diffraction, angular spectrum method or by other reconstruction approaches. There is no need for calibration experiments, and no *a priori* knowledge about the object is required. Both, lateral and axial resolution of the reconstructed object is greatly enhanced after three-dimensional volumetric deconvolution is applied.

**OCIS codes:** (090.1995) Digital holography; (100.1830) Deconvolution; (180.6900) Three-dimensional microscopy.

Keywords: digital holography, volumetric deconvolution, three-dimensional volumetric deconvolution, particle tracking, holographic particle tracking, resolution, PSF

# Contents

## 1. Introduction

The beauty of holography lies within the possibility to restore three-dimensional information from a two-dimensional interference pattern. In digital holography (*1-3*), the widely used scientific application of holography, the hologram is usually formed as an interference image captured by a CCD camera and has no physical z-thickness. As a result, the three-dimensional distribution of the object wave-front reconstructed from such a digital hologram suffers from a limited depth of focus and the reconstructed objects do not appear well localized but are prolonged in z-direction. Some discussion on depth-resolution in digital holography can be found in (*4-6*), including recently proposed method of depth-resolution improvement by intensity averaging (*6*). In the process of reconstruction of digital holograms, the three-dimensional object wave-front is obtained as a set of two-dimensional complex distributions at various distances from the hologram plane, which is analogous to 'optical sectioning' in optical microscopy. In every such two-dimensional section there is signal from the in-focus part of the object but also blurred signal from the out-of-focus part of the object. Consequently, the object wave-front reconstructed from a hologram does not directly represent the original object's distribution. Here we present methods for retrieving the depth-resolved original object's distribution from its hologram.

## 2. Image formation as convolution with the point-spread function

The problem of out-of focus signal has been studied in optical microscopy and has a number of solutions by applying the so-called deconvolution methods; overviews of the methods can be found in (*7-10*). When images of a three-dimensional object $O(\vec{r})$ are acquired at several focal planes (optical sections), a three-dimensional volume of measured data $M(\vec{r})$ is created:

$$O(\vec{r}) \Rightarrow M(\vec{r})\,. \qquad (1)$$

The image of a point scatterer through the same optical system forms the point-spread function (PSF) of the system:

$$\delta(\vec{r}) \Rightarrow \mathrm{PSF}(\vec{r})\,. \qquad (2)$$

In optical microscopy, the PSF is usually recorded experimentally by using a small object, e.g. a microsphere.

If the object can be represented as a sum of point scatterers (condition of linearity):

$$O(\vec{r}) = \int O(\vec{s})\delta(\vec{r}-\vec{s})\mathrm{d}\vec{s} \qquad (3)$$

and the optical system images the individual scatterers in the same way (condition of shift invariance), the resulting image can be described as the sum of images of the individual scatterers (*11*):

$$M(\vec{r}) = \int O(\vec{s})\mathrm{PSF}(\vec{r}-\vec{s})\mathrm{d}\vec{s}\,. \qquad (4)$$

Equation 4 can be viewed as a convolution:

$$M(\vec{r}) = O(\vec{r}) \otimes \mathrm{PSF}(\vec{r})\,. \qquad (5)$$

The deconvolution of the measured three-dimensional data volume $M(\vec{r})$ with the known PSF of the system makes it possible to obtain the three-dimensional object $O(\vec{r})$ .

## 3. Formation and reconstruction of a hologram

An important difference between imaging in a conventional microscope and holography is that coherence and interference are the key elements for the latter. The interference between waves scattered by different parts of the object must be accounted for. The wave-front reconstructed from a hologram carries amplitude and phase information and exhibits a complex distribution. The requirement of linearity is fulfilled for the complex optical fields: the total complex optical field created by a set of scatterers is the sum of the complex optical fields created by each scatterer. An additional requirement in the case of holography is that the effects of multiple scattering between different parts of the object must be negligible. To fulfill the requirement of shift invariance we consider plane waves for simplicity (in case of a spherical wave-front geometrical scaling must be applied in addition).

When an incident plane wave is scattered by some three-dimensional object $O(\vec{r})$, neglecting multiple scattering effects, the resultant object wave-front at any point $\vec{r}$ can be represented as a convolution of the object function with the impulse response of free space propagation:

$$U_{\text{Object}}(\vec{r}) = \iiint O(\vec{r}_{\text{O}}) \frac{\exp(ik|\vec{r}_{\text{O}} - \vec{r}|)}{|\vec{r}_{\text{O}} - \vec{r}|} \mathrm{d}\vec{r}_{\text{O}} = O(\vec{r}) \otimes \frac{\exp(ikr)}{r} \qquad (6)$$

where the integration is performed over the object's coordinates $\vec{r}_{\text{O}}$.

The superposition of the wave scattered by the object $U_{\text{Object}}(\vec{r}_{\text{S}})$ and a reference wave $U_{\text{R}}(\vec{r}_{\text{S}})$, where $\vec{r}_{\text{S}} = (x_{\text{S}}, y_{\text{S}}, z_{\text{S}})$, when recorded at some distance $z_{\text{D}}$ produces a hologram, see Fig. 1.

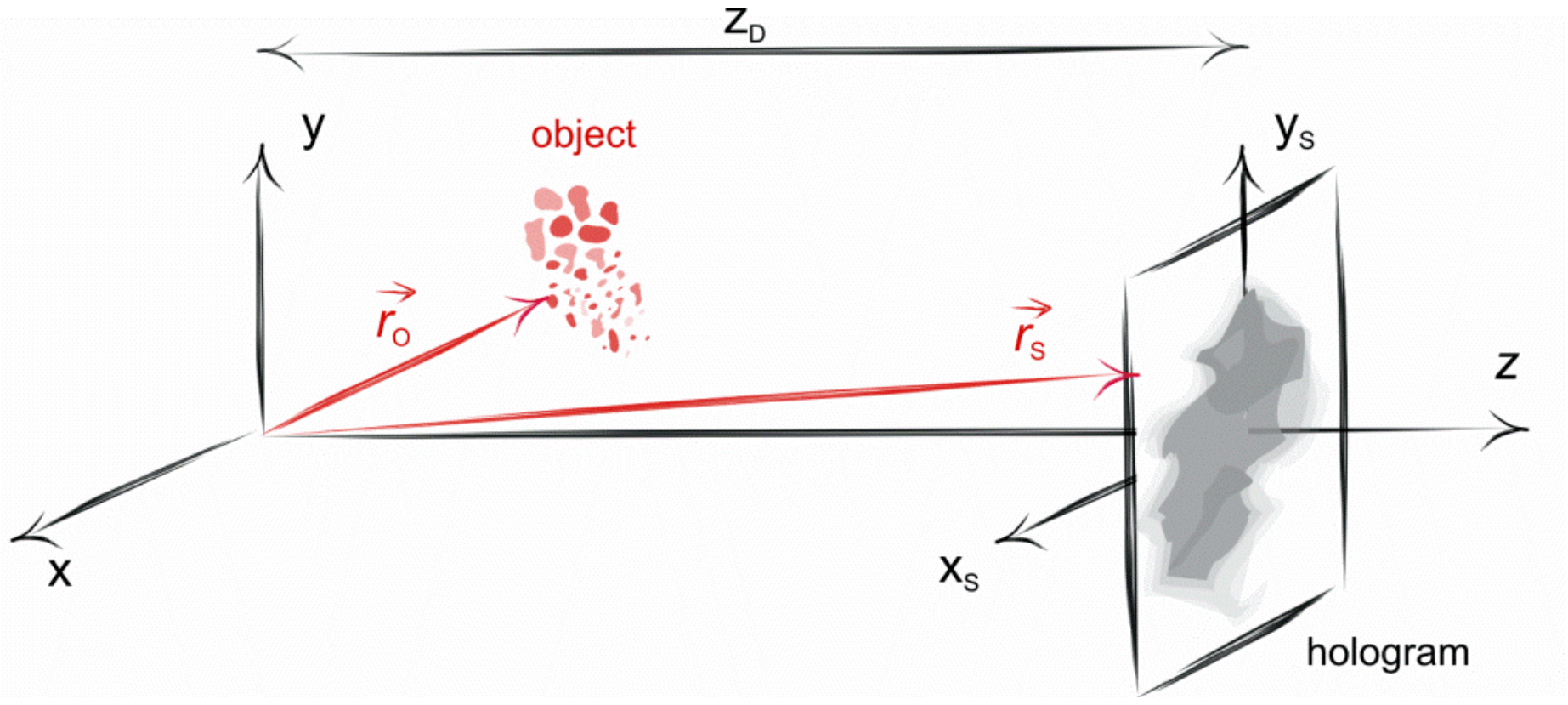

Fig.1. Scheme of hologram recording and illustration to the used symbols.

The distribution of the transmission function of the recorded hologram is given by $\tilde{H}(\vec{r}_{\text{S}}) = \left|\tilde{U}_{\text{Object}}(\vec{r}_{\text{S}}) + \tilde{U}_{\text{R}}(\vec{r}_{\text{S}})\right|^2$, where the tilde denotes the two-dimensionality, i.e. the function is defined only in the screen $(x_{\text{S}}, y_{\text{S}})$-plane. After subtraction of the constant background $\left|\tilde{U}_{\text{R}}(\vec{r}_{\text{S}})\right|^2$ and in the approximation of a weak object wave $\left|\tilde{U}_{\text{Object}}(\vec{r}_{\text{S}})\right|^2 << \left|\tilde{U}_{\text{R}}(\vec{r}_{\text{S}})\right|^2$, the distribution in the hologram plane is given by:

$$\tilde{H}(\vec{r}_{\text{S}}) \sim \tilde{U}_{\text{Object}}(\vec{r}_{\text{S}})\tilde{U}^*_{\text{R}}(\vec{r}_{\text{S}}) + \tilde{U}^*_{\text{Object}}(\vec{r}_{\text{S}})\tilde{U}_{\text{R}}(\vec{r}_{\text{S}})\,. \qquad (7)$$

The mathematical routine for the reconstruction of holograms is independent of the nature of the recorded objects and consists of the following two steps:

1. Illumination with the reference wave and

2. Backward propagation to the object-position.

In a first step, the multiplication of the hologram with the simulated complex reference wave $\tilde{U}_{\mathrm{R}}(\vec{r}_{\mathrm{S}})$, provides the fully restored *two-dimensional* complex distribution of the object wave in the hologram plane: $\tilde{U}_{\mathrm{R}}(\vec{r}_{\mathrm{S}})\tilde{H}(\vec{r}_{\mathrm{S}}) \sim \tilde{U}_{\mathrm{Object}}(\vec{r}_{\mathrm{S}})$. In a second step, the backward propagation of the complex wave-front from the hologram plane to the object's location is calculated by applying the Huygens-Fresnel principle:

$$U_{\mathrm{O}}(\vec{r}) = \iiint \tilde{U}_{\mathrm{Object}}(\vec{r}_{\mathrm{S}})\delta(z_{\mathrm{S}} - z_{\mathrm{D}}) \frac{\exp(ik|\vec{r}_{\mathrm{S}} - \vec{r}|)}{|\vec{r}_{\mathrm{S}} - \vec{r}|} \mathrm{d}\vec{r}_{\mathrm{S}} \qquad (8)$$

and the result of the integration provides the reconstructed *three-dimensional* wave-front of the object wave.

In analogy to optical microscopy, the wave-front reconstructed from the hologram of a point-scatterer shall serve as a basis for the PSF:

$$U_{\mathrm{P}}(\vec{r}) = \iiint \tilde{U}_{\mathrm{Point}}(\vec{r}_{\mathrm{S}})\delta(z_{\mathrm{S}} - z_{\mathrm{D}}) \frac{\exp(ik|\vec{r}_{\mathrm{S}} - \vec{r}|)}{|\vec{r}_{\mathrm{S}} - \vec{r}|} \mathrm{d}\vec{r}_{\mathrm{S}}\,. \qquad (9)$$

Eq. (8) and (9) are usually referred to in literature as reconstruction by convolution with the impulse response of free space propagation (see for instance (*12*)). Next, we would like to demonstrate, how the object distribution can be extracted from the reconstructed object wave-front by employing a three-dimensional volumetric deconvolution with the point-spread function. For the three-dimensional volumetric deconvolution procedure there are two three-dimensional distributions of the reconstructed wave-fronts available: $U_{\mathrm{O}}(\vec{r})$ and $U_{\mathrm{P}}(\vec{r})$. The method of three-dimensional volumetric deconvolution is applied to the reconstructed wavefronts. It is universal, as it does not depend on how the reconstructed wavefronts were obtained, by employing Huygens-Fresnel diffraction, Rayleigh-Sommerfeld diffraction, angular spectrum method or other reconstruction approaches.

## 4. Three-dimensional volumetric deconvolution of the reconstructed optical fields

In direct analogy to optical microscope data, the complex optical field reconstructed from a hologram of the object $U_{\mathrm{O}}(\vec{r})$ can be considered as the measured signal $M(\vec{r})$ and the complex optical field reconstructed from a hologram of a point scatterer as the $\mathrm{PSF}(\vec{r})$. We thus obtain the convolution equation, described by Eq. (5), as:

$$U_{\mathrm{O}}(\vec{r}) = O(\vec{r}) \otimes U_{\mathrm{P}}(\vec{r})\,, \qquad (10)$$

where $O(\vec{r})$ is the unknown three-dimensional object distribution.

The simplest way to solve Eq. (10) is to use the convolution theorem (FT here refers to a three-dimensional Fourier transform):

$$\mathrm{FT}(U_{\mathrm{O}}(\vec{r})) = \mathrm{FT}\left(O(\vec{r})\right)\mathrm{FT}(U_{\mathrm{P}}(\vec{r}))\,. \qquad (11)$$

and solve the equation by division:

$$O(\vec{r}) = \mathrm{FT}^{-1}\left(\frac{\mathrm{FT}(U_{\mathrm{O}}(\vec{r}))}{\mathrm{FT}(U_{\mathrm{P}}(\vec{r}))}\right). \qquad (12)$$

However, such a direct approach will not result in the object's three-dimensional distribution for the following reason. The reconstructed optical fields given by Eq. (8) and (9) can be represented as a three-dimensional convolution with the impulse response of the free space propagation and the convolution theorem can be applied:

$$\mathrm{FT}(U_{\mathrm{O}}(\vec{r}))=\mathrm{FT}(\tilde{U}_{\mathrm{Object}}(\vec{r})\delta(z-z_{\mathrm{D}}))\mathrm{FT}\left(\frac{\exp(ikr)}{r}\right) \tag{13}$$

and:

$$\mathrm{FT}(U_{\mathrm{P}}(\vec{r}))=\mathrm{FT}(\tilde{U}_{\mathrm{Point}}(\vec{r})\delta(z-z_{\mathrm{D}}))\mathrm{FT}\left(\frac{\exp(ikr)}{r}\right), \tag{14}$$

where the three-dimensional Fourier transform is given by:

$$\mathrm{FT}(U(\vec{r}))=\iiint U(\vec{r})\exp(-2\pi i\vec{r}\vec{q})\mathrm{d}\vec{r} \tag{15}$$

or, in Cartesian coordinates:

$$\mathrm{FT}(U(x,y,z))=\iiint U(x,y,z)\exp(-2\pi i(x\mu+y\nu+z\eta))\mathrm{d}x\mathrm{d}y\mathrm{d}z \tag{16}$$

where $\vec{q}=(\mu,\nu,\eta)$ are coordinates in the Fourier domain.

Next, substitution of Eq. (13) and Eq. (14) into Eq. (12) results in elimination of the three-dimensional Fourier transforms of the free-space propagators, and only a fraction of the three-dimensional Fourier transforms of the two-dimensional images remains:

$$O(\vec{r})=\mathrm{FT}^{-1}\left(\frac{\mathrm{FT}(\tilde{U}_{\mathrm{Object}}(\vec{r})\delta(z-z_{\mathrm{D}}))}{\mathrm{FT}(\tilde{U}_{\mathrm{Point}}(\vec{r})\delta(z-z_{\mathrm{D}}))}\right), \tag{17}$$

which after integration over the *z*-coordinate are in detail

$$\mathrm{FT}(\tilde{U}_{\mathrm{Object}}(\vec{r})\delta(z-z_{\mathrm{D}}))=\exp(-2\pi iz_{\mathrm{D}}\eta)\iint\tilde{U}_{\mathrm{Object}}(x,y)\exp(-2\pi i(x\mu+y\nu))\mathrm{d}x\mathrm{d}y$$

$$\mathrm{FT}(\tilde{U}_{\mathrm{Point}}(\vec{r})\delta(z-z_{\mathrm{D}}))=\exp(-2\pi iz_{\mathrm{D}}\eta)\iint\tilde{U}_{\mathrm{Point}}(x,y)\exp(-2\pi i(x\mu+y\nu))\mathrm{d}x\mathrm{d}y\,. \tag{18}$$

By substituting the last two equations into Eq. (17), we obtain:

$$O(\vec{r})=\mathrm{FT}^{-1}\left(\frac{\iint\tilde{U}_{\mathrm{Object}}(x,y)\exp(-2\pi i(x\mu+y\nu))\mathrm{d}x\mathrm{d}y}{\iint\tilde{U}_{\mathrm{Point}}(x,y)\exp(-2\pi i(x\mu+y\nu))\mathrm{d}x\mathrm{d}y}\right). \tag{19}$$

Here, obviously, the fraction in the brackets does not have a dependency on $\eta$ and thus, the backward three-dimensional Fourier transform will always result in a $\delta$-function at $z=0$ as a result of integration over $\mathrm{d}\eta$. As a consequence, the object function $O(\vec{r})$ will always be just a two-dimensional distribution in the plane $z=0$. This implies that such direct three-dimensional volumetric deconvolution applied to the reconstructed complex fields would always result in losing the $z$-information.

We suggest two ways to solve the problem:
1. A three-dimensional volumetric deconvolution of the *intensities* of the reconstructed fields.
2. An iterative three-dimensional volumetric deconvolution.

## 4.1 Instant three-dimensional volumetric deconvolution of the reconstructed intensities

One way to perform an instant three-dimensional volumetric deconvolution is to apply it to the reconstructed *intensity* distributions instead of the complex amplitudes. Mathematically, when intensities are considered instead of complex amplitudes, it allows avoiding elimination of $z$ and η -dependent terms while performing the division given by Eq. (12). In fact, all deconvolution methods employed in optical microscopy are designed to be applied to the intensity distributions (*7-10*). We consider the reconstructed intensity of the object wave $\left|U_{\mathrm{O}}(\vec{r})\right|^2 = I_{\mathrm{O}}(\vec{r})$ as the measured signal $M(\vec{r})$ and the reconstructed intensity of a point scatterer $\left|U_{\mathrm{P}}(\vec{r})\right|^2 = I_{\mathrm{P}}(\vec{r})$ as the $\mathrm{PSF}(\vec{r})$. In the case of intensities, the requirement of linearity is only fulfilled when the effects of interference between waves from individual point scatterers constituting the object are negligible. This condition, however, is not unrealistic, but very common for instance in the holographic study of particle fields where a large amount of identical particles is spread in a volume (*13-14*). The instant three-dimensional volumetric deconvolution is calculated as follows:

$$O(\vec{r}) = \mathrm{FT}^{-1}\left(\frac{\mathrm{FT}\left[I_{\mathrm{O}}(\vec{r})\right]}{\mathrm{FT}\left[I_{\mathrm{P}}(\vec{r})\right]}\right) \approx \mathrm{FT}^{-1}\left(\frac{\mathrm{FT}\left[I_{\mathrm{O}}(\vec{r})\right]\left\{\mathrm{FT}\left[I_{\mathrm{P}}(\vec{r})\right]\right\}^*}{\left|\mathrm{FT}\left[I_{\mathrm{P}}(\vec{r})\right]\right|^2+\beta}\right), \tag{20}$$

where $\beta$ is a small addend to avoid division by zero; $\beta$ is usually much smaller than $\left|\mathrm{FT}\left[I_{\mathrm{P}}(\vec{r})\right]\right|^2$ and selected to achieve a good signal-to-noise reconstruction. The result of the three-dimensional volumetric deconvolution is a set of sharp localizations (ideally, $\delta$-functions) of individual scatterers constituting the object.

## 4.2 Iterative deconvolution of the reconstructed complex fields

The method of iterative three-dimensional volumetric deconvolution solves Eq. (11) without direct division. The following example shall be considered to illustrate that. Two point scatterers are given: point scatterer A shall be positioned at $\vec{r}_{\mathrm{A}} = (0,0,0)$ and scatterer B at $\vec{r}_{\mathrm{B}} = (x_{\mathrm{B}}, y_{\mathrm{B}}, z_{\mathrm{B}})$. The scattered complex wave-fronts are described by:

$$U_{\mathrm{A}}(\vec{r}_{\mathrm{S}}) \approx \frac{\exp\left(ik\sqrt{x_S{}^2+y_S{}^2+z_S{}^2}\right)}{z_{\mathrm{S}}} \tag{21}$$

and

$$U_{\mathrm{B}}(\vec{r}_{\mathrm{S}}) \approx \frac{\exp\left(ik\sqrt{(x_S-x_{\mathrm{B}})^2+(y_S-y_{\mathrm{B}})^2+(z_S-z_{\mathrm{B}})^2}\right)}{(z_{\mathrm{S}}-z_{\mathrm{B}})}. \tag{22}$$

The two holograms of the scatterers are recorded at $z_{\mathrm{S}} = z_{\mathrm{D}}$. The complex field reconstructed from the hologram of scatterer A is employed as the PSF of the optical system. The complex field reconstructed from the hologram of scatterer B is the reconstructed object wave-front. When Eq. (21) and Eq. (22) are substituted into Eq. (18), we obtain the following three-dimensional distribution in the Fourier domain:

$$\mathrm{FT}(\tilde{U}_{\mathrm{A}}(\vec{r})\delta(z-z_{\mathrm{D}})) = \exp(-2\pi i z_{\mathrm{D}}\eta)\exp(-\pi i\lambda z_{\mathrm{D}}(\mu^2+\nu^2)) \tag{23}$$

$$\mathrm{FT}(\tilde{U}_{\mathrm{B}}(\vec{r})\delta(z - z_{\mathrm{D}})) =$$
$$= \exp(-2\pi i z_{\mathrm{D}}\eta)\exp(-2\pi i(x_{\mathrm{B}}\mu + y_{\mathrm{B}}\nu))\exp(-\pi i\lambda(z_{\mathrm{D}} - z_{\mathrm{B}})(\mu^2 + \nu^2)) . \quad (24)$$

Obviously, when substituted in Eq. (12), the direct division of Eq. (24) with Eq. (23) will lead to the elimination of the $\eta$-dependency and consequently the loss of z-information. When Eq. (24) and (23) are substituted into the convolution equation (11), by taking into account the three-dimensional Fourier transform of the impulse response of free space propagator:

$$\mathrm{FT}\left(\frac{\exp(ikr)}{r}\right) \approx \delta\left(\eta + \frac{\lambda}{2}(\mu^2 + \nu^2)\right), \quad (25)$$

after some simplifications, we arrive at Eq. (11) as:

$$\exp(-2\pi i(x_{\mathrm{B}}\mu + y_{\mathrm{B}}\nu))\exp(\pi i\lambda z_{\mathrm{B}}(\mu^2 + \nu^2))\delta\left(\eta + \frac{\lambda}{2}(\mu^2 + \nu^2)\right) =$$
$$= \mathrm{FT}\left(O(\vec{r})\right)\delta\left(\eta + \frac{\lambda}{2}(\mu^2 + \nu^2)\right). \quad (26)$$

Equation (26) can be solved at $\eta = -\frac{\lambda}{2}(\mu^2 + \nu^2)$ and its solution is:

$$\mathrm{FT}\left(O(\vec{r})\right) = \exp\left(-2\pi i(x_{\mathrm{B}}\mu + y_{\mathrm{B}}\nu + z_{\mathrm{B}}\eta)\right). \quad (27)$$

From this, the exact position of scatterer B is obtained by a three-dimensional backward Fourier transformation. A direct division could thus be avoided and Eq. (11) is solved by carefully matching the left and right sides which reveals the recovery of the object distribution. In the next section we verify this theoretical hypothesis by applying it to simulated and experimental examples. Since this iterative deconvolution is applied to the complex optical fields rather than to the intensities, it can also be applied to continuous extended objects where the interference between different parts of the object cannot be neglected anymore.

The iterative loop is adapted from optical microscopy and includes the following steps (*15*):

$$O^{(1)}(\vec{r}) = U_{\mathrm{O}}(\vec{r})$$

(i) $\quad U_{\mathrm{O}}^{(\mathrm{k})}(\vec{r}) = O^{(\mathrm{k})}(\vec{r}) \otimes U_{\mathrm{P}}(\vec{r})$ $\quad$ (28)

(ii) $\quad O^{(\mathrm{k}+1)}(\vec{r}) = O^{(\mathrm{k})}(\vec{r})\frac{U_{\mathrm{O}}(\vec{r})}{U_{\mathrm{O}}^{(\mathrm{k})}(\vec{r})}$

(iii) $\quad \mathrm{k} = \mathrm{k} + 1$

The division in step (ii) is calculated as

$$O^{(\mathrm{k}+1)}(\vec{r}) = O^{(\mathrm{k})}(\vec{r})\frac{U_{\mathrm{O}}(\vec{r})\left(U_{\mathrm{O}}^{(\mathrm{k})}(\vec{r})\right)^*}{\left|U_{\mathrm{O}}^{(\mathrm{k})}(\vec{r})\right|^2 + \beta} \quad (29)$$

where $\beta$ is a small addend to avoid division by zero. The selection criteria for $\beta$ have to fulfill the following conditions: $\beta$ has to be much smaller than the maximal value of $\left|U_{\mathrm{O}}^{(\mathrm{k})}(\vec{r})\right|^2$ but sufficiently large to ensure convergence of the iterative process. With each iteration step, the object function $O^{(\mathrm{k})}(\vec{r})$ and the reconstructed object wave $U^{(\mathrm{k})}{}_{\mathrm{O}}(\vec{r})$ are approaching the complex distribution of the true object function $O(\vec{r})$ and the reconstructed wave-front $U_O(\vec{r})$, respectively, while $U_P(\vec{r})$ remains unaltered. The three-dimensional distributions obtained at each iterative step can be multiplied with the apodization function to avoid accumulation of artifacts during repeated three-dimensional Fourier-transformations. The apodization function selected here consists of a three-dimensional spherical filter of radius R, which leaves all pixel values inside the sphere unchanged and sets the ones outside the sphere to zero.

## 5. Numerical simulation and reconstruction of holograms

Before applying the routine to simulated and experimental examples, we would like to describe in some detail how the numerical optical field propagation used in the simulation and reconstruction routines is carried out. For the calculation of the complex field propagation we used the well-known algorithm for plane wave propagation (see for example (*16*)) that consists of the following three steps:

(a) Two-dimensional forward Fourier-transform of the complex field at the plane $z_i$.

(b) Multiplication with the simulated complex transfer function: $\exp\left(\frac{2\pi i}{\lambda} z\sqrt{1-\mu^2-\nu^2}\right)$

where $z = z_{i+1} - z_i$ is the propagation distance, $\mu = \frac{\lambda}{S} p, \nu = \frac{\lambda}{S} q$, where $S$ is the area size, $p = -\frac{N}{2} \ldots \frac{N}{2}$ and $q = -\frac{N}{2} \ldots \frac{N}{2}$ are the pixel indexes, and $N$ is the number of pixels.

(c) Two-dimensional backward Fourier-transform of the product (a) and (b).

The reconstruction procedure consists of the same steps (a) – (c), but the complex transfer function is the complex conjugated one due to the back propagation: $z \to -z$.

A PSF was created as the field distribution reconstructed from a hologram of a point scatterer. The hologram of a point scatterer could be either experimentally recorded or simulated numerically. The hologram for the PSF must be of the same size as the hologram of the object and it should be reconstructed for the same z-span. The result of three-dimensional volumetric deconvolution shows the relative shift of the object's parts to the position of the point scatterer.

All images were sampled with 200 × 200 pixels size. The reconstructions were done at a set of *z*-distances for 200 steps in total. The resulting three-dimensional data volume exhibited a size of 200 × 200 × 200 pixels limited by the ability of the employed MATLAB software to operate with arrays of larger size. To calculate the three-dimensional FFT we used the MATLAB function "fftn", which allows FFT of any dimension, together with "fftshift" and "ifftshift" commands to keep the lower frequencies in the center of the complex images. In the iterative reconstruction, a single iteration takes about 2– 5 seconds.

## 6. Three-dimensional volumetric deconvolution of simulated holograms

To simulate a three-dimensional object, we used a set of two-dimensional transmission functions at various *z*-distances. The complex optical field was consequently propagated between each pair of *xy*-planes. Thus, multiple scattering effects were taken into account to generate a simulated hologram as realistic as possible to check whether multiple scattering effects have an influence on the three-dimensional volumetric deconvolution.

### 6.1 Instant three-dimensional volumetric deconvolution of the reconstructed intensities

For the instant three-dimensional volumetric deconvolution the condition of negligible interference between wave-fields scattered from different parts constituting the object must be satisfied. We thus simulated a particle distribution in the shape of the three Greek letters α, β and γ, each made up of point scatterers and placed at three planes at different distances from the screen; see Fig. 2(a).

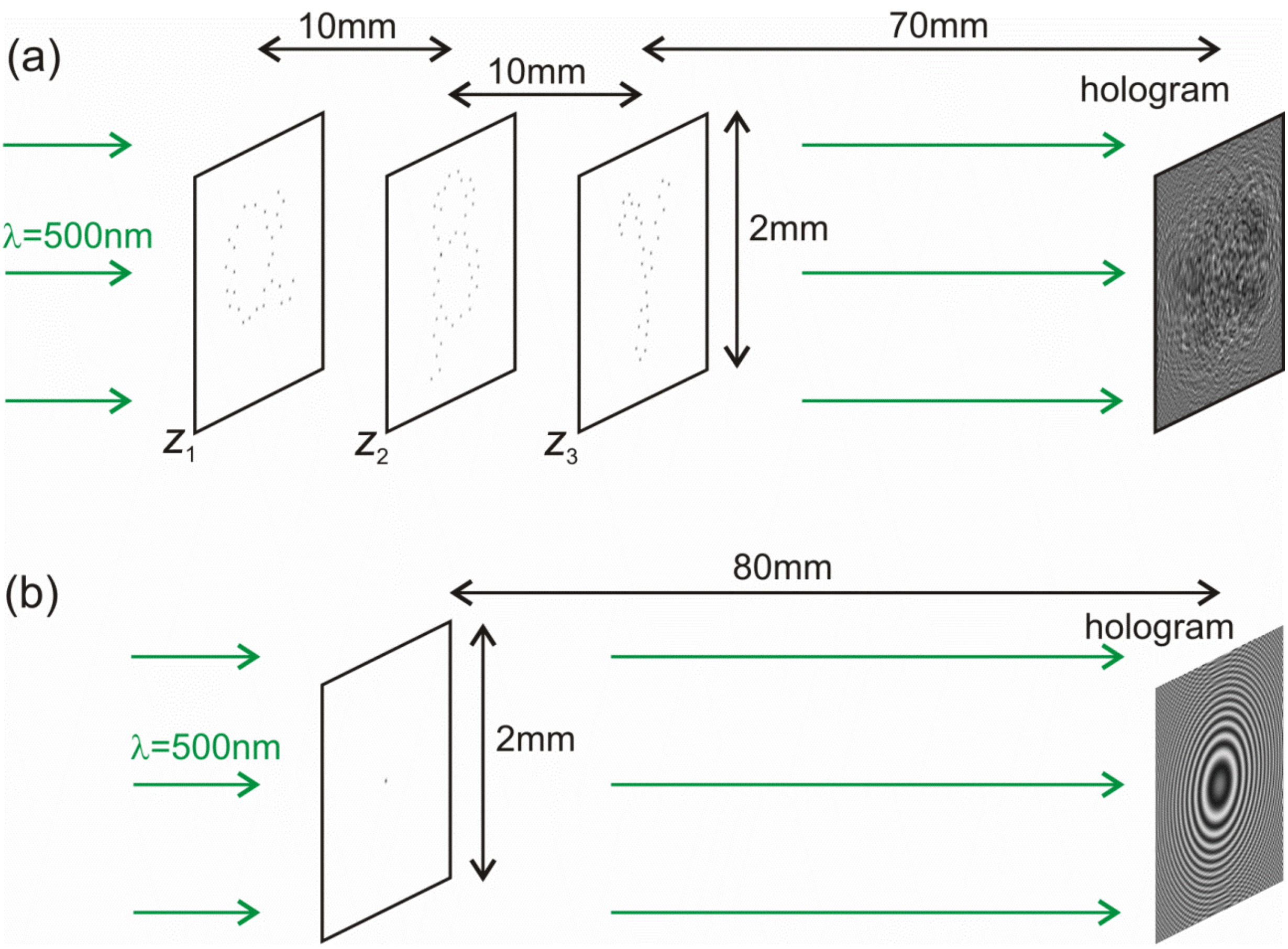

Fig.2. Arrangement of numerical experiments for the simulation of holograms of a particle distribution and of an individual point scatterer. (a) Simulation of the hologram of a particle distribution showing the transmission functions in the three planes at which the particles are arranged. (b) Simulation of the hologram of a point scatterer.

The letters α, β and γ were positioned at a 90, 80 and 70 mm distance from the screen plane, respectively. The incident radiation exhibited a wavelength of 500 nm. Object and the screen area sizes were selected to be 2 mm. The reconstructions were carried out from z = 40 to 120 mm distance with a step-width of 0.4 mm for a number of 200 steps in total. The PSF was created as the reconstruction of the hologram of a point scatterer positioned at 80 mm from the screen; see Fig. 2(b).

The three-dimensional amplitude distribution reconstructed from the simulated hologram shows the presence of the in-focus and out-of-focus signal, see Fig. 3. The instant three-dimensional volumetric deconvolution was performed using Eq. (18) with $\beta = 1$. The three-dimensional volumetric deconvolution leads to the removal of the out-of-focus signal. The individual scatterers revealed sharply and the patterns of the three Greek letters could clearly be distinguished, see Fig. 3 and Fig. 4. The *z*-localization of a selected particle in the deconvoluted distribution amounts to less than 1.6 mm in the reconstruction volume distributed over 80 mm in z-direction, see Fig.4(d). In addition, any noise caused by the out-of focus signals and twin images was removed.

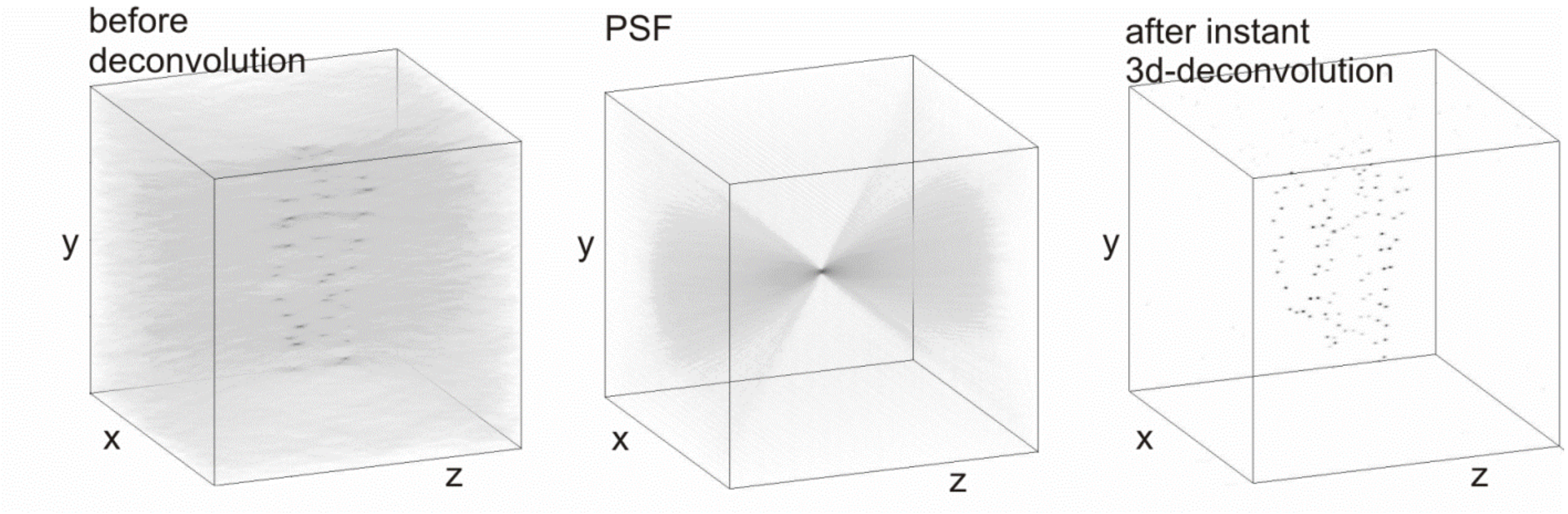

Fig.3. (a) Three-dimensional representation of the amplitude of the reconstructed complex wave-front. (b) Amplitude of the complex PSF. (c) Result of the instant three-dimensional volumetric deconvolution.

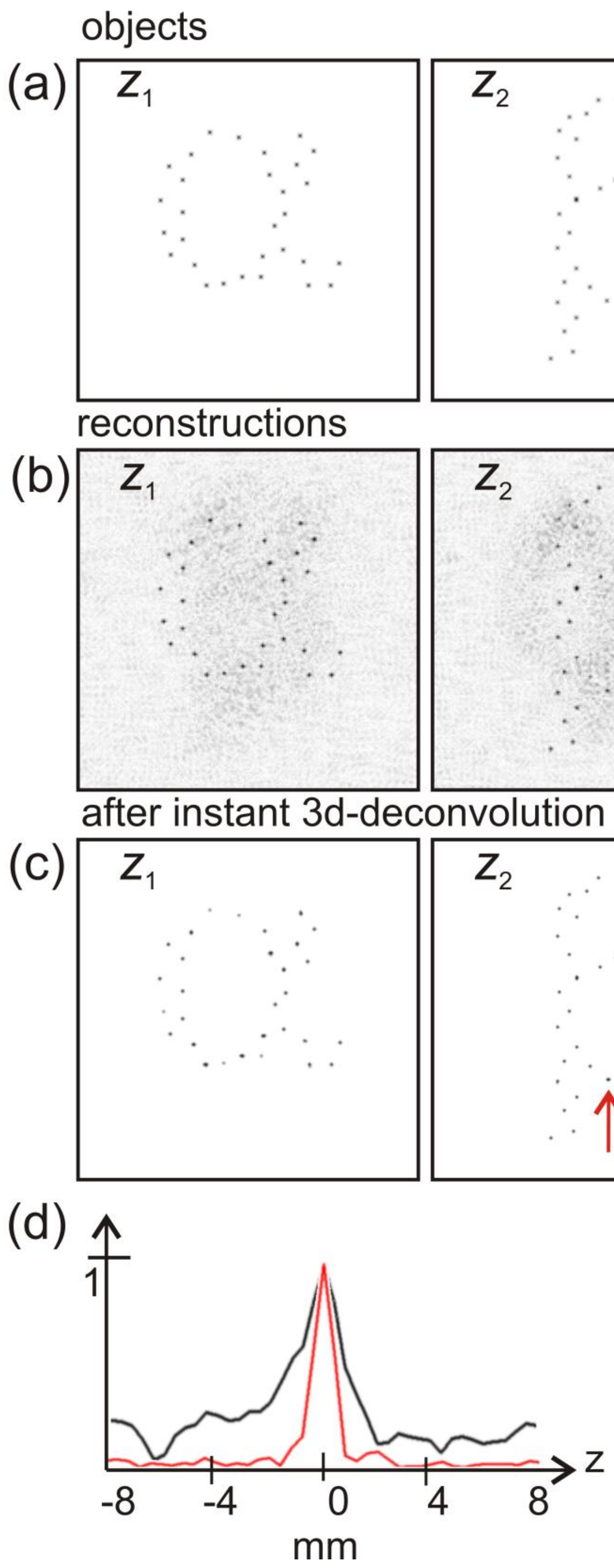

Fig.4. Results of the instant three-dimensional volumetric deconvolution of the intensities reconstructed from the simulated hologram of a particle distribution. (a) Original objects used in the simulation at the three $z$-planes. (b) Reconstructed amplitude of the object wave-front at the same $z$-planes. (c) Result of instant three-dimensional volumetric deconvolution. (d) Intensity profiles along $z$-direction for the particle indicated by the red arrow in (c) before (black) and after (red) instant three-dimensional volumetric deconvolution.

## 6.2 Iterative three-dimensional volumetric deconvolution of the reconstructed complex field

The iterative three-dimensional volumetric deconvolution was applied to the same simulated hologram of the particle distribution using Eq. (28-29) with $\beta = 0.01$. After each step (i) of the iterative loop, the absolute value of the function $U_O^{(k)}(\vec{r})$ was normalized so that its minimum and maximum match the absolute value of $U_O(\vec{r})$. The result after 24 iterations is shown in Fig. 5. The *z*-localization of a selected particle amounts to just 1.6 mm in the deconvoluted object distribution, as shown in Fig. 5 (b). The error function, computed as:

$$\mathrm{Error}(k) = \frac{\sum_{i,j,l} \left| \left| U_O^{(k)}(i,j,l) \right| - \left| U_O(i,j,l) \right| \right|}{\sum_{i,j,l} \left| U_O(i,j,l) \right|} \tag{30}$$

shows a decrease after the first iterations followed by oscillations around some minimal value. However, the error function does not actually reach zero. This is because the updated complex field $U^{(k)}{}_O(\vec{r})$ which is simulated as the convolution of the updated object distribution function $O^{(k)}(\vec{r})$ with the PSF does not include such effects as multiple scattering or twin images and is thus never exactly equal to the reconstructed complex field $U_O(\vec{r})$.

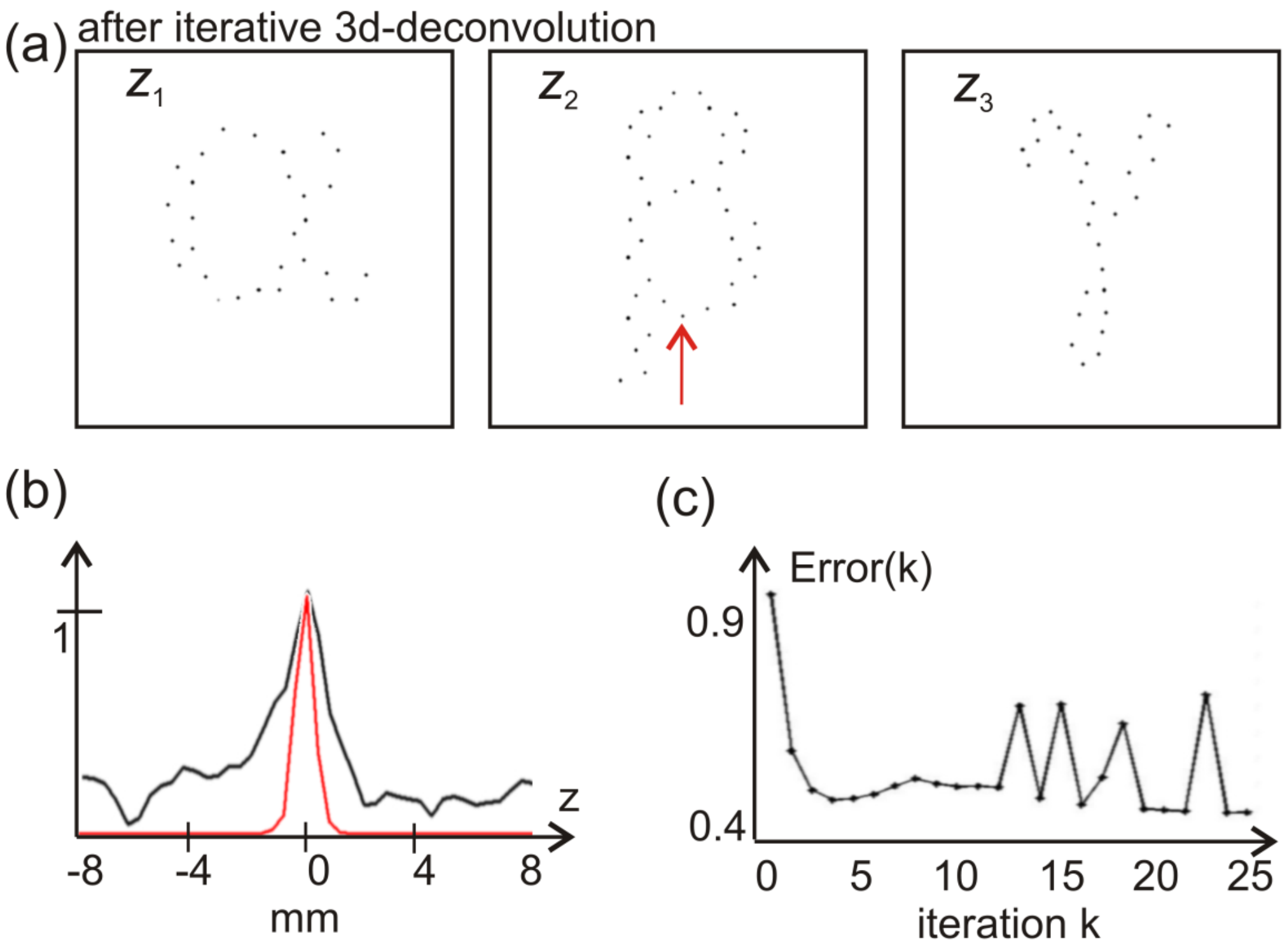

Fig.5. (a) Results of the iterative three-dimensional volumetric deconvolution of the complex fields reconstructed from the simulated hologram of a particle distribution. (a) Amplitude distributions of the deconvoluted fields at the three planes. (b) Intensity profiles for a 16 mm span around the in-focus position along z-direction for the particle indicated by an arrow in (a) before and after deconvolution. (c) Error as a function of the iteration number.

The iterative three-dimensional volumetric deconvolution can also be applied to restore continuous, non-point-like objects. As an example, we simulated a hologram of the three continuous Greek letters: $\alpha$, $\beta$ and $\gamma$ placed at a 90, 80 and 70 mm distance from the screen. The other parameters of the hologram simulation and reconstruction remained the same as in the example described above. The three-dimensional volumetric deconvolution was done with the iterative loop Eq. (28-29) with $\beta = 0.1$. The function $O^{(k+1)}(\vec{r})$ obtained after every step (ii) in the iterative loop was multiplied with the apodization function - a three-dimensional spherical filter of radius $R$ = 80 pixels. To smooth the appearance of continuous objects and avoid divergence of the routine caused by random peaks of noise, a Gaussian low-pass filter was applied:

$$\mathrm{GLPF}(\mu,\nu,\eta) = \exp\left(-\frac{\mu^2+\nu^2+\eta^2}{2D^2}\right), \tag{31}$$

where $\mu,\nu,\eta$ are coordinates in the Fourier-domain and $D$ is a small constant. In our example, the filter was applied in every 5$^{th}$ iteration with $D$=5. The results of the iterative three-dimensional volumetric deconvolution are shown in Fig. 6.

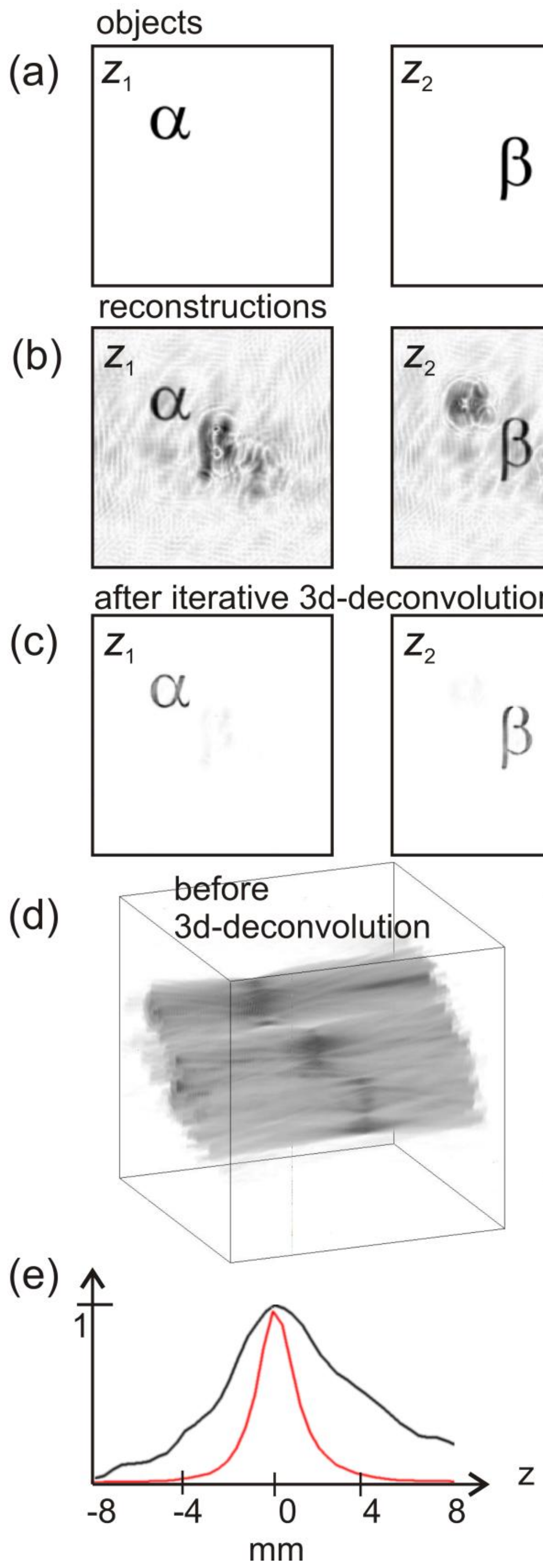

Fig.6. (a) Results of the iterative three-dimensional volumetric deconvolution of the complex fields reconstructed from the simulated hologram of continuous objects. (a) Original objects used in the simulation at the three different planes. (b) Reconstructed objects at the same planes. (c) Result of the three-dimensional volumetric deconvolution. (d) Three-dimensional representation of the reconstructed amplitude of the object wave-front and amplitude of the deconvoluted object function. (e) Intensity profiles for a 16mm span around the in-focus position along z-direction for the point of the γ-object indicated by the red arrow in (c) before (black) and after (red) iterative three-dimensional volumetric deconvolution.

## 7. Experimental optical holograms

To apply our method also to experimental data, we recorded optical holograms of a three-dimensional object: a piece of a 170 μm thin glass plate with polystyrene microspheres of 4 μm in diameter randomly distributed on both sides of the glass (S1 and S2, see Fig. 7). The employed holographic inline scheme using plane waves is shown in Fig. 7. The wavelength of the green laser employed amounts to 532 nm, the distance between object and microscope objective was about 1 mm, the imaged area about 550 μm in diameter, the distance between the microscope objective and the screen was selected to 16 cm, a hologram of about 30 mm in diameter was recorded on a semitransparent screen and captured by a CCD camera.

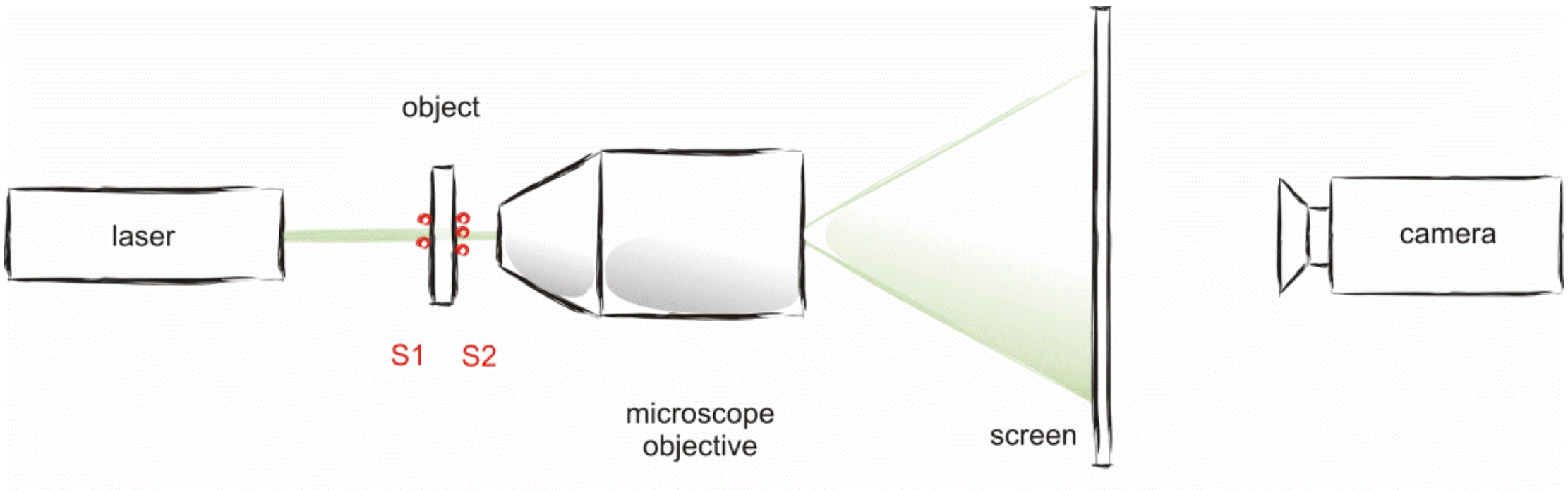

Fig.7. Experimental scheme for in inline holographic recording with plane waves.

Directly after the acquisition of the hologram H0 of the sample, an image in the absence of the sample was recorded – the background image B. Thereafter the laser beam was blocked by a piece of black cardboard and the intensity of the dark environment – IDE – was recorded. The IDE amounts to about 50 gray levels on the 10 Bit (1024 gray levels) CCD camera chip. The normalization of the hologram was calculated by the division H=(H0-IDE)/(B-IDE). The result was a contrast hologram H of average intensity of 1; an example of which is shown in Fig. 8.

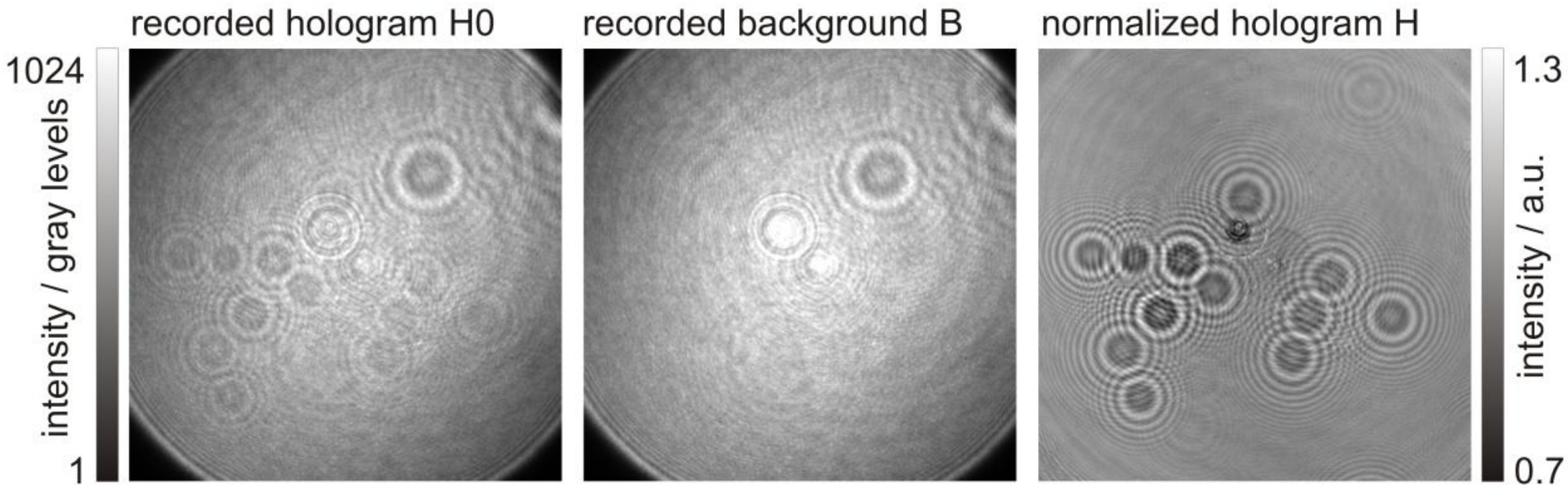

Fig.8. Experimentally recorded hologram H0 of polystyrene spheres (left), the background image B (center) and the normalized hologram H (right).

Polystyrene exhibits an absorption coefficient of about 3 $cm^{-1}$ (*17*) and a refractive index of about 1.6 at the wavelength of 532 nm. A 4μm diameter polystyrene sphere thus constitutes a weak absorbing but a strong (up to 28 radians) phase object. Below we show that even such strong phase shifting properties represent no problem for our deconvolution routine.
The recorded holograms exhibited an original size of 1000 × 1000 pixels but needed to be re-sampled to 200 × 200 pixels for the reconstruction and three-dimensional volumetric deconvolution process for reasons explained in Section 5. Re-sampling was done using a linear interpolation in both, *x*- and *y*-direction, for computing pixel locations and their gray values. The information about individual scatterers is stored in the interference fringes which are distributed over the entire hologram surface. Thus, re-sampling of the hologram does not alter the information about the position of the scatterers but just introduces a loss of spatial resolution in the reconstructions.

### 7.1 Instant three-dimensional volumetric deconvolution of the reconstructed intensities

The optical holograms of the spheres were normalized and reconstructed, see Fig. 9. The reconstructions were carried out from z = 170 to 1770 μm distance with a step-width of 8 μm for a number of 200 steps in total. The two planes of in-focus spheres were found at distances of 970 μm (S1) and 800 μm (S2) from the hologram plane, as shown in Fig. 9(d). The hologram for the PSF was created as a cut-out from a hologram of the spheres where just an individual sphere was detected, see Fig. 9(a), and the cut-out was subsequently placed into the image's center and zero-padded, as shown in Fig. 9(b). The instant three-dimensional volumetric deconvolution using Eq.(18) with $\beta = 0.01$ was applied to the reconstructed intensity distribution. As a result, the well-localized *z*-positions of the objects were retrieved free from out-of focus-signals, as shown in Fig. 9(e).

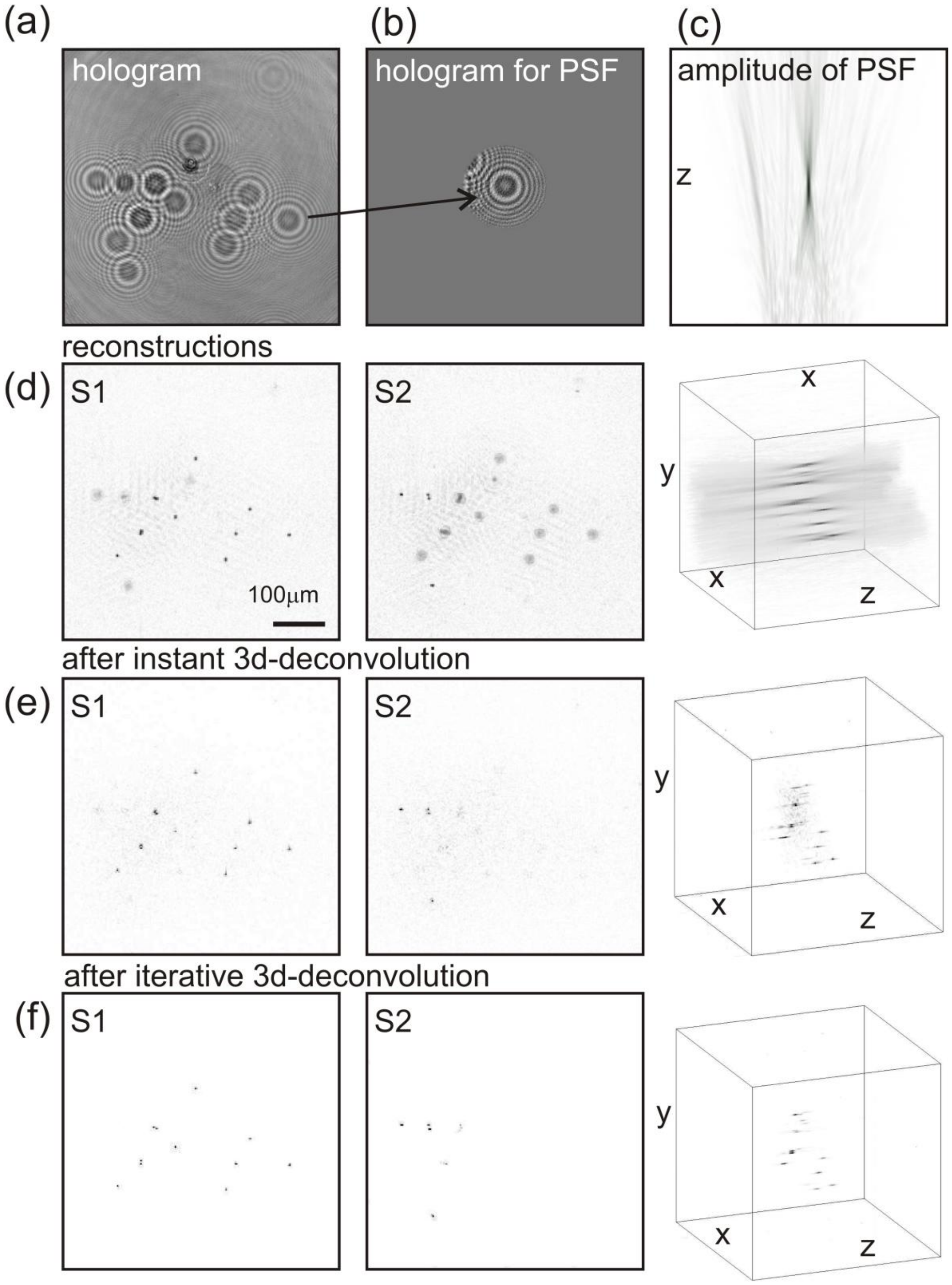

Fig.9. three-dimensional volumetric deconvolution of experimental holographic reconstruction. (a) Normalized hologram. (b) Hologram for the PSF as a cut out of the experimental hologram related to just a single sphere. (c) Amplitude reconstruction of the hologram for the PSF, slice in *xz*-plane. (d) Amplitude reconstruction at both sides of the microscopic slide, at the S1 and S2 plane. In-focus and out-of of focus spheres can be distinguished. Right: three-dimensional representation of the reconstructed amplitude distribution. (e) Results after instant three-dimensional volumetric deconvolution. Amplitude distributions at the S1 and S2 plane together with the three-dimensional distribution. (f) Results after iterative three-dimensional volumetric deconvolution. Amplitude distributions at the S1 and S2 plane and three-dimensional distribution.

### 7.2 Iterative three-dimensional volumetric deconvolution of the reconstructed complex fields

For the iterative three-dimensional volumetric deconvolution we used the complex distribution of the optical field reconstructed from the same experimental hologram as in the example above. The iterative three-dimensional volumetric deconvolution was done using 12 iterations by applying Eq. (28-29) with $\beta = 0.012$. The function $O^{(k+1)}(\vec{r})$ obtained after every step (ii) in the iterative loop was multiplied with the apodization function - a three-dimensional spherical filter of radius $R$ = 98 pixels. No other numerical filters were applied. The results are shown in Fig. 9(f) and exhibit superior removal of the out-of-focus signal when compared to that of the instant three-dimensional volumetric deconvolution, as evident from comparing Fig. 9(e) with Fig. 9(f).

Apparently, the three-dimensional volumetric deconvolution also improves the resolution in the *xy*-plane, as illustrated in Figure 10 where we show the magnified images of the reconstructions at the plane S1 before and after the iterative three-dimensional volumetric deconvolution. While the clusters of spheres are originally not well resolved, after applying the three-dimensional volumetric deconvolution scheme, the positions of the individual spheres composing the clusters are clearly visible and emerge as individual sharp spots. Both, lateral and axial resolution of the reconstructed object is greatly enhanced. After applying three-dimensional deconvolution the individual spheres that are 4 μm apart are resolved.

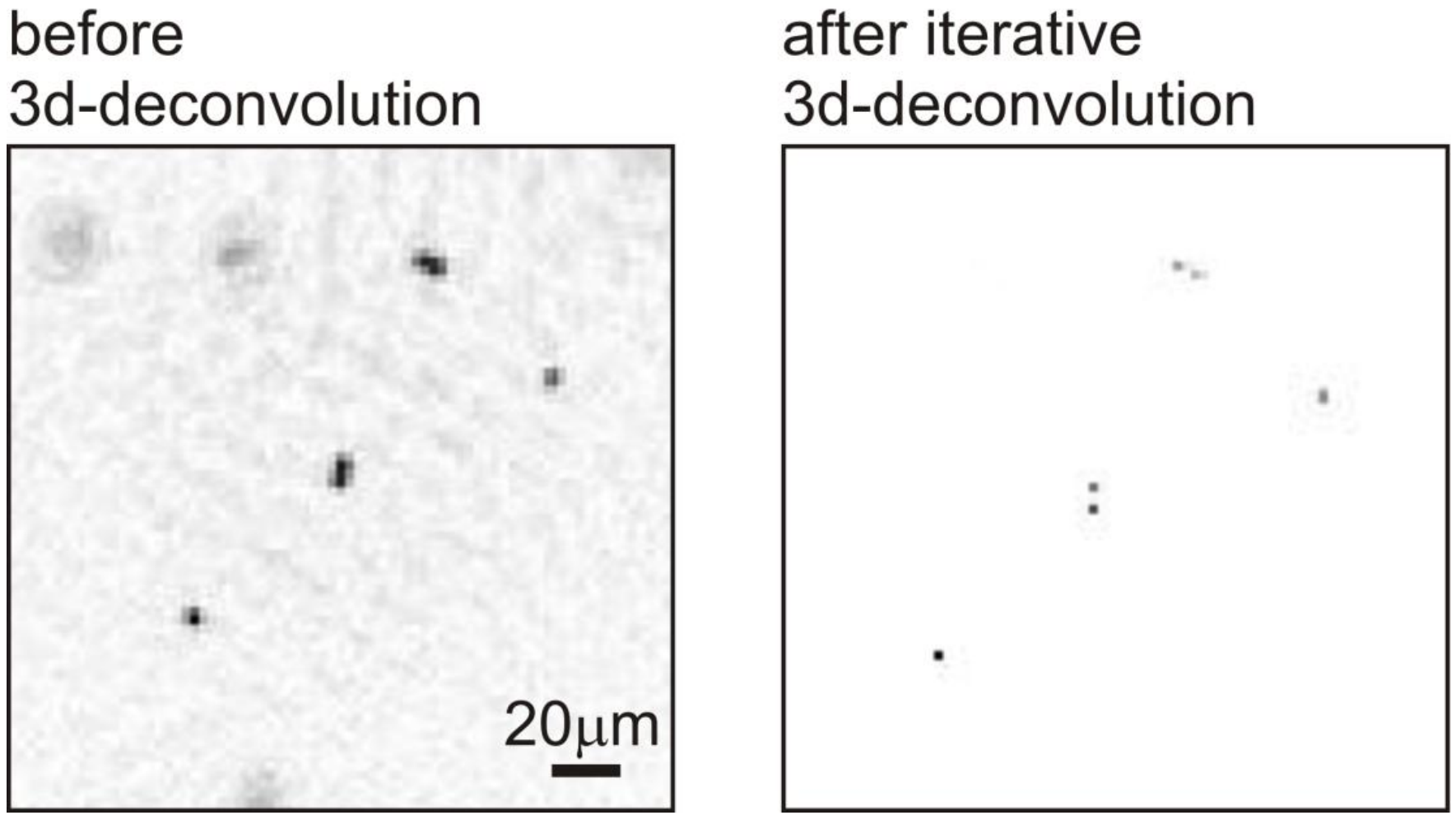

Fig.10. Magnified images of the reconstructed amplitude before and after iterative three-dimensional volumetric deconvolution at plane S1.

### 7.3 Three-dimensional volumetric deconvolution of the reconstructed optical fields with a simulated PSF

In some cases, the PSF is not available from experimental data. For instance, when a hologram does not contain an image of an individual stand-alone sphere. In this case the hologram of a point scatterer can be simulated and its complex reconstruction provides the PSF. To check this approach, we simulated a hologram of a point scatterer in the middle of a 550 × 550 μm$^2$ area placed at 970 μm distance from the screen, as shown in Fig. 11(a) and we used the reconstruction of its hologram as the PSF for three-dimensional volumetric deconvolution. For the instant three-dimensional volumetric deconvolution we selected $\beta$ =

0.01 and for the iterative three-dimensional volumetric deconvolution $\beta = 0.2$. The results of the instant and iterative three-dimensional volumetric deconvolutions are shown in Fig. 11(b) and (c), respectively. The resulting distributions of spheres are similar to the distributions obtained by three-dimensional volumetric deconvolution with the experimental PSF, shown in Fig. 9 (e) and (f). The instant three-dimensional volumetric deconvolution using the simulated PSF, as shown in Fig. 11(b), results in point-like reconstructed objects, however with some remaining background noise left. The iterative three-dimensional volumetric deconvolution with the simulated PSF provides noise-free reconstructed spheres, however they appear slightly extended in *z*-direction. The result demonstrates that comparably good results of three-dimensional volumetric deconvolution can be obtained using either an experimentally obtained PSF or a simulated one.

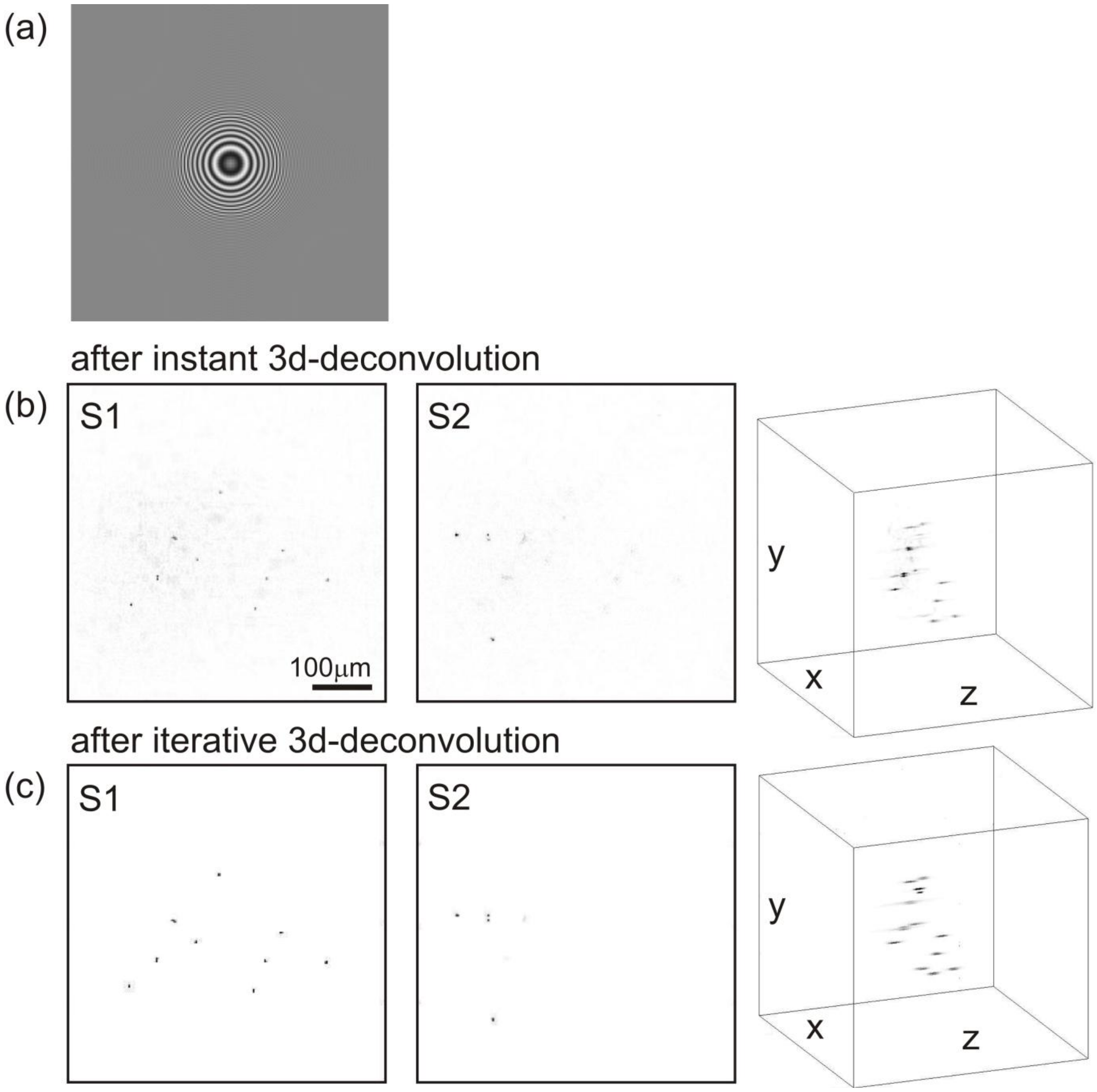

Fig.11. Three-dimensional volumetric deconvolution of the experimental holographic reconstruction with a simulated PSF. (a) Simulated hologram of a point scatterer. (b) Results of instant three-dimensional volumetric deconvolution at S1 and S2 planes and as three-dimensional representation. (b) Results of iterative three-dimensional volumetric deconvolution at S1 and S2 planes and as three-dimensional representation.

## 7.4 Iterative three-dimensional volumetric deconvolution of the reconstructed complex fields applied to extended objects

In this section we show that the method of the iterative three-dimensional volumetric deconvolution with the simulated PSF can also be applied to complex and/or not-point-like objects, such as two spheres of a extended size. A 200 × 200 pixel area corresponding to 110 × 110 μm$^2$ was selected in an experimental hologram of spheres; see Fig. 12(a). The reconstructions were carried out from z = 10 to 1610 μm distance with a step-width of 8μm for a number of 200 steps in total. The amplitude reconstruction shows two spheres at different z-distances: at 810 μm (S1) and 630 μm (S2) from the hologram. The PSF was obtained by the reconstruction of the complex field distribution from the simulated hologram of a point scatterer placed in the middle of 110 × 110 μm$^2$ area at the distance of $z$ = 810 μm from the screen. The iterative three-dimensional volumetric deconvolution was performed with 12 iterations using $\beta$ = 0.012 without any numerical filters. The results are shown in Fig. 12(b-d): after the iterative three-dimensional volumetric deconvolution the shape of the spheres appears sharper in all three directions and the out-of focus signal is removed.

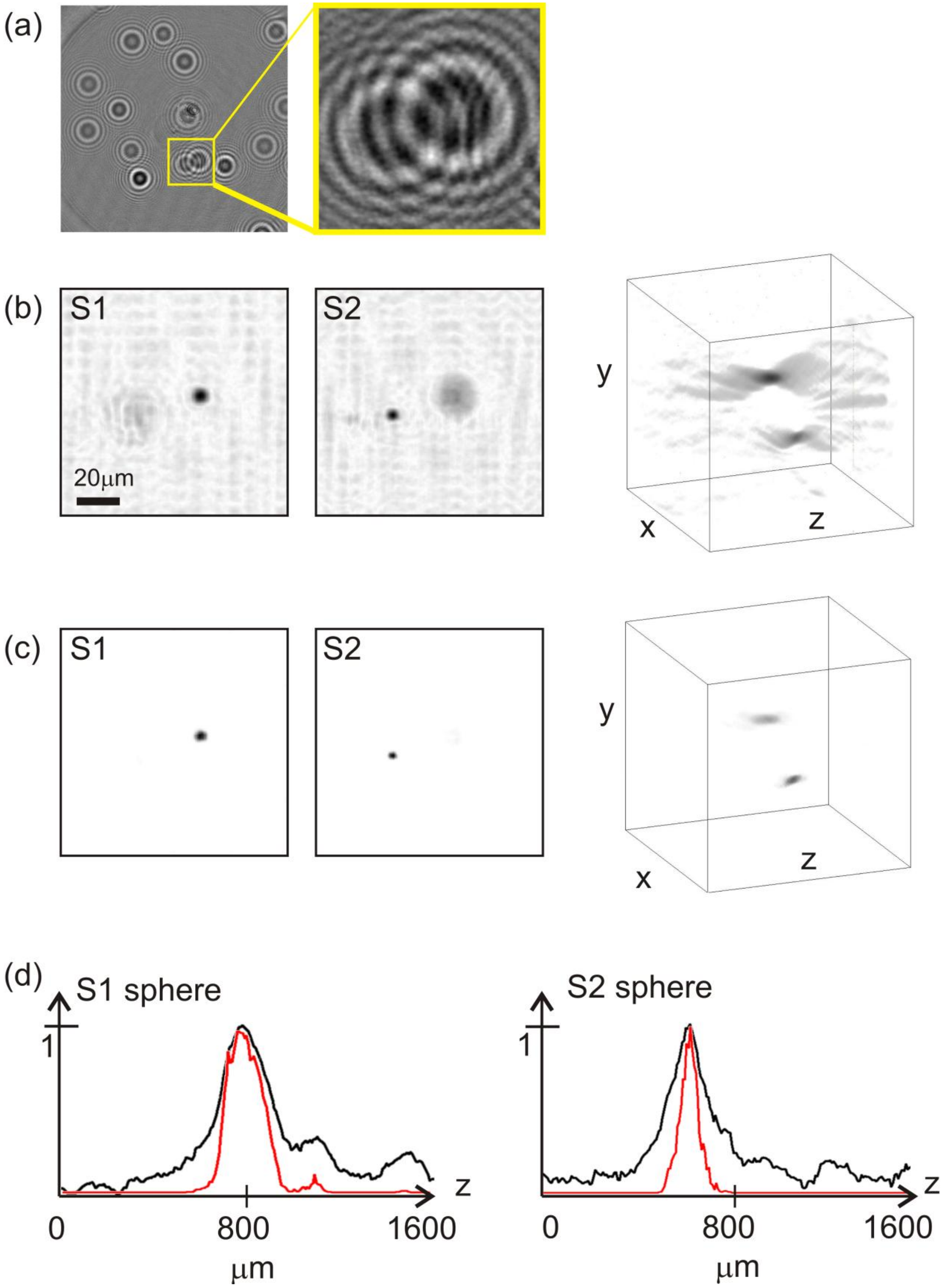

Fig. 12. Results of iterative three-dimensional volumetric deconvolution of the optical field reconstructed from an experimental hologram with a simulated PSF. (a) Selected fragment of an experimental hologram of spheres. (b) Reconstructed amplitude distribution at planes S1 and S2 and its three-dimensional representation. (c) Results of the iterative three-dimensional volumetric deconvolution at the plane S1 and S2 together with its three-dimensional representations. (d) Amplitude profiles along z-direction at the spheres' positions before (black) and after (red) three-dimensional volumetric deconvolution.

## 8. Conclusions

We have shown that three-dimensional volumetric deconvolution methods as known from optical microscopy can also be applied to restore true object distribution from holographic records and their reconstructions. We demonstrated methods of instant and iterative three-dimensional volumetric deconvolutions applied to simulated as well as to experimental holograms. The three-dimensional distribution of polystyrene spheres of 4 μm in diameter deposited on glass surfaces could be retrieved free from disturbing out-of focus and twin image signals by a three-dimensional volumetric deconvolution of the hologram reconstructions. Both, lateral and axial resolution of the reconstructed object is greatly enhanced. In the experimental holograms, after three-dimensional deconvolution the individual spheres that are 4 μm apart were resolved.

Instant three-dimensional volumetric deconvolution can be applied if the object consists of a set of spaced-apart particles distributed in a certain volume. The method of instant three-dimensional volumetric deconvolution has potential to be applied in holographic particle tracking velocimetry, where the identification of the correct *z*-position of the object and the removal of the out-of-focus signal is essential. The iterative three-dimensional volumetric deconvolution can also be applied for objects exhibiting complex and/or extended shapes or distributions. In view of future applications, it is important to note that the methods of three-dimensional volumetric deconvolution are independent of the nature of the coherent radiation used. This opens up the possibility for three-dimensional reconstruction also in X-ray or electron holography. The method of three-dimensional volumetric deconvolution is universal, it does not depend on how the reconstructed wavefronts were obtained, by employing Huygens-Fresnel diffraction, Rayleigh-Sommerfeld diffraction, angular spectrum method or other reconstruction approaches. There is no need for calibration experiments, and no *a priori* knowledge about the object is required. The PSF required for three-dimensional volumetric deconvolutions can either be obtained by experimentally recording and reconstructing a hologram of an arbitrarily small individual scatterer or as the reconstruction of a simulated hologram of a point scatterer, which in general is similar to a Fresnel zone-plate. The approach of creating the PSF from a simulated hologram can also be applied to account for angular-dependent scattering, as it is the case of low-energy electron scattering. A realistic but simulated PSF can thus be used in electron holography where an experimental hologram of a point scatterer is not readily available.

By employing three-dimensional volumetric deconvolution methods the out-of focus signal is brought back to its scatterer and the twin images are automatically removed as they are not part of the scattered wave. Thus, spatially well localized parts of an object are recovered free from artifacts and the beauty of holography as a three-dimensional imaging technique lives up to its full potential.

## 9. Acknowledgements

The authors are grateful to Conrad Escher for sample preparation and to Daniel Eaton for careful reading of the manuscript. Financial support of the European Project SIBMAR and the Forschungskredit of the University of Zurich are gratefully acknowledged.

# 10. MATLAB code

## 10.1 README

"a_simulate_PSF" simulates hologram of a points source and saves it as BIN and JPG
"b_reconstruction_and_deconvolution.m" reconstructs b_hologram.bin and makes deconvolution of the reconstructed intensities.
The other programs are subroutines and can be placed into your „C: .... MATLAB“ folder
%%%%%%%%%%%%%%%%%%%%%%%%%%%%%%%%%%%%%%%%%%%%

parameters:
N = 200 pixels
wavelength = 500nm
area size = 2mm x 2mm
object z-span: z = 400um - 1200um
point scatterer is located in the middle of z-span at z = 800um

## 10.2 Simulation of hologram of point scatterer

Save it as “a_simulate_PSF.m”

```
%%%%%%%%%%%%%%%%%%%%%%%%%%%%%%%%%%%%%%%%%%%%%%%%%%%%%%%%%%%%%%%%%%%%%%%%%%
% SIMULATION OF HOLOGRAM OF POINT SCATTERER
%%%%%%%%%%%%%%%%%%%%%%%%%%%%%%%%%%%%%%%%%%%%%%%%%%%%%%%%%%%%%%%%%%%%%%%%%%
% Citation for this code/algorithm or any of its parts:
% Tatiana Latychevskaia, Fabian Gehri and Hans-Werner Fink
% "Depth-resolved holographic reconstructions by three-dimensional
deconvolution",
% Optics Express 18(21), 22527 - 22544 (2010)
%%%%%%%%%%%%%%%%%%%%%%%%%%%%%%%%%%%%%%%%%%%%%%%%%%%%%%%%%%%%%%%%%%%%%%%%%%
% The code is written by Tatiana Latychevskaia, 2010
% tatiana(at)physik.uzh.ch
% The version of Matlab for this code is R2010b

clear all
close all

addpath('C:/Program Files/MATLAB/R2010b/myfiles');
%%%%%%%%%%%%%%%%%%%%%%%%%%%%%%%%%%%%%%%%%%%%%%%%%%%%%%%%%%%%%%%%%%%%%%%%%%
% PARAMETERS

N = 200;                    % number of pixels
lambda = 500*10^(-9);       % wavelength in meter
area = 0.002;               % area size in meter
z = 0.08;                   % z distance in meter
%%%%%%%%%%%%%%%%%%%%%%%%%%%%%%%%%%%%%%%%%%%%%%%%%%%%%%%%%%%%%%%%%%%%%%%%%%
% CREATING SMOOTHED POINT SCATTERER
object = zeros(N,N);
reference = ones(N,N);

object(N/2-1:N/2+1,N/2-1:N/2+1) = 0.25;
object(N/2,N/2) = 1;

figure, imshow(rot90(object), []);
%%%%%%%%%%%%%%%%%%%%%%%%%%%%%%%%%%%%%%%%%%%%%%%%%%%%%%%%%%%%%%%%%%%%%%%%%%
% SIMULATING HOLOGRAM OF THE POINT SCATTERER

prop = Propagator(N, lambda, area, z);
Uobj = IFT2Dc(FT2Dc(-object).*conj(prop));
Uref = IFT2Dc(FT2Dc(reference).*conj(prop));

hologram = abs(Uobj + Uref).^2;
%%%%%%%%%%%%%%%%%%%%%%%%%%%%%%%%%%%%%%%%%%%%%%%%%%%%%%%%%%%%%%%%%%%%%%%%%%
% SAVING HOLOGRAM OF THE POINT SCATTERER

fid = fopen(strcat('a_point_source_hologram.bin'), 'w');
fwrite(fid, hologram, 'real*4');
fclose(fid);

figure, imshow(rot90(hologram), []);
colormap(gray)
p = hologram;
p = 255*(p - min(min(p)))/(max(max(p)) - min(min(p)));
imwrite (rot90(p), gray, 'a_point_source_hologram.jpg');
%%%%%%%%%%%%%%%%%%%%%%%%%%%%%%%%%%%%%%%%%%%%%%%%%%%%%%%%%%%%%%%%%%%%%%%%%%
```

## 10.3 Reconstruction of plane wave holograms and 3d-deconvolution, volumetric deconvolution

Save it as “b_reconstruction_and_deconvolution.m”

```
%%%%%%%%%%%%%%%%%%%%%%%%%%%%%%%%%%%%%%%%%%%%%%%%%%%%%%%%%%%%%%%%%%%%%%%%%%%
% RECONSTRUCTION OF PLANE WAVE HOLOGRAMS
% AND 3D-DECONVOLUTION, VOLUEMTRIC DECONVOLUTION
%%%%%%%%%%%%%%%%%%%%%%%%%%%%%%%%%%%%%%%%%%%%%%%%%%%%%%%%%%%%%%%%%%%%%%%%%%%
% Citation for this code/algorithm or any of its parts:
% Tatiana Latychevskaia, Fabian Gehri and Hans-Werner Fink
% "Depth-resolved holographic reconstructions by three-dimensional
deconvolution",
% Optics Express 18(21), 22527 - 22544 (2010)
%%%%%%%%%%%%%%%%%%%%%%%%%%%%%%%%%%%%%%%%%%%%%%%%%%%%%%%%%%%%%%%%%%%%%%%%%%%
% The code is written by Tatiana Latychevskaia, 2010
% tatiana(at)physik.uzh.ch
% The version of Matlab for this code is R2010b

clear all
close all

% addpath('C:/Program Files/MATLAB/R2010b/myfiles');
%%%%%%%%%%%%%%%%%%%%%%%%%%%%%%%%%%%%%%%%%%%%%%%%%%%%%%%%%%%%%%%%%%%%%%%%%%%
% PARAMETERS

N = 200;                      % number of pixels
lambda = 500*10^(-9);         % wavelength in meter
area = 0.002;                 % area size in meter
z_start = 0.04;               % z distance start in meter
z_end = 0.12;                 % z distance end in meter
z_step = 0.0004;              % z distance step in meter
beta = 1;                     % Parameter beta in intensities deconvolution

S = round((z_end - z_start)/z_step);

%%%%%%%%%%%%%%%%%%%%%%%%%%%%%%%%%%%%%%%%%%%%%%%%%%%%%%%%%%%%%%%%%%%%%%%%%%%
% READING HOLOGRAM FOR PSF

    fid = fopen('a_point_source_hologram.bin', 'r');
    hologramPSF = fread(fid, [N, N], 'real*4');
    fclose(fid);
    hologramPSF = hologramPSF - 1;

%%%%%%%%%%%%%%%%%%%%%%%%%%%%%%%%%%%%%%%%%%%%%%%%%%%%%%%%%%%%%%%%%%%%%%%%%%%
% RECONSTRUCTION OF HOLOGRAM FOR PSF

reconstructionPSF = zeros(N,N,S);
for ii = 1:S
z = z_start + ii*z_step;
prop = Propagator(N, lambda, area, z);
recPSF = IFT2Dc(FT2Dc(hologramPSF).*prop);
PSF(:,:,ii) = recPSF(:,:);
end

PSF = permute(PSF, [2 1 3]);
%%%%%%%%%%%%%%%%%%%%%%%%%%%%%%%%%%%%%%%%%%%%%%%%%%%%%%%%%%%%%%%%%%%%%%%%%%%
% 3D PLOT OF PSF
```

```
    figure, axes('fontsize', 12, 'xtick', 0:floor(N/4):N, 'ytick',
0:floor(N/4):N, 'ztick', 0:floor(N/4):N)
    vol3d('CData', sqrt(abs(permute(PSF, [2 3 1]))), 'texture', '3D')
    view(-202, 18), axis tight
    colormap('gray'), colormap(1 - colormap), colorbar
    xlabel('z', 'fontsize', 16)
    ylabel('x', 'fontsize', 16)
    zlabel('y', 'fontsize', 16)
    axis([0 N 0 N 0 N])
    box on, zoom(0.7)
    alphamap('decrease', 0)

%%%%%%%%%%%%%%%%%%%%%%%%%%%%%%%%%%%%%%%%%%%%%%%%%%%%%%%%%%%%%%%%%%%%%%%%%%%%
% READING HOLOGRAM OF THE OBJECT

    fid = fopen('b_hologram.bin', 'r');
    hologramO = fread(fid, [N, N], 'real*4');
    fclose(fid);
    hologramO = hologramO - 1;

%%%%%%%%%%%%%%%%%%%%%%%%%%%%%%%%%%%%%%%%%%%%%%%%%%%%%%%%%%%%%%%%%%%%%%%%%%%%
% RECONSTRUCTION OF THE OBJECT WAVEFRONT

reconstructionO = zeros(N,N,S);

for ii=1:S
z = z_start + ii*z_step;
prop = Propagator(N, lambda, area, z);
recO = IFT2Dc(FT2Dc(hologramO).*prop);
reconstructionO(:,:,ii) = recO(:,:);
end

reconstructionO = permute(reconstructionO, [2 1 3]);

%%%%%%%%%%%%%%%%%%%%%%%%%%%%%%%%%%%%%%%%%%%%%%%%%%%%%%%%%%%%%%%%%%%%%%%%%%%%
% 3D PLOT OF RECONSTRUCTED OBJECT WAVEFRONT

    figure, axes('fontsize', 12, 'xtick', 0:floor(N/4):N, 'ytick',
0:floor(N/4):N, 'ztick', 0:floor(N/4):N)
    vol3d('CData', abs(permute(reconstructionO, [2 3 1])), 'texture', '3D')
    view(-202, 18), axis tight
    colormap('gray'), colormap(1 - colormap), colorbar
    xlabel('z', 'fontsize', 16)
    ylabel('x', 'fontsize', 16)
    zlabel('y', 'fontsize', 16)
    axis([0 N 0 N 0 N])
    box on, zoom(0.7)
    alphamap('decrease', 0.05)

%%%%%%%%%%%%%%%%%%%%%%%%%%%%%%%%%%%%%%%%%%%%%%%%%%%%%%%%%%%%%%%%%%%%%%%%%%%%
% 3D-DECONVOLUTION (VOLUEMTRIC DECONVOLUTION) OF INTENSITIES

% intensities
reconstruction_int = reconstructionO .* conj(reconstructionO);
PSF_int = PSF .* conj(PSF);

% Reconstruction = Fourier transform of reconstructed intensity
Reconstruction = FT3Dc(reconstruction_int);
clear reconstruction_int
```

```
% CTF = Fourier transform of PSF
CTF = FT3Dc(PSF_int);
clear PSF_int

%%%%%%%%%%%%%%%%%%%%%%%%%%%%%%%%%%%%%%%%%%%%%%%%%%%%%%%%%%%%%%%%%%%%%%%%%%%%%%
CTFi = conj(CTF) ./ (CTF .* conj(CTF) + beta);
clear CTF

Object = Reconstruction .* CTFi;
clear Reconstruction, clear CTFi

object = IFT3Dc(Object);
clear Object

%%%%%%%%%%%%%%%%%%%%%%%%%%%%%%%%%%%%%%%%%%%%%%%%%%%%%%%%%%%%%%%%%%%%%%%%%%%%%%
% PLOTS OF OBJECT

    figure, axes('fontsize', 12, 'xtick', 0:floor(N/4):N, 'ytick',
0:floor(N/4):N, 'ztick', 0:floor(N/4):N)
    vol3d('CData', abs(permute(object, [2 3 1])), 'texture', '3D')
    view(-202, 18), axis tight
    colormap('gray'), colormap(1 - colormap), colorbar
    xlabel('z', 'fontsize', 16)
    ylabel('x', 'fontsize', 16)
    zlabel('y', 'fontsize', 16)
    axis([0 N 0 N 0 N])
    box on, zoom(0.7)
    alphamap('decrease', 0.05)

zcut = 25; % z-position of xy-cut out (counted from the center z)
           % zcut = 0 for "beta"
           % zcut = -25 for "gamma"
           % zcut = 25 for "alpha"
    object_xy = object(:, :, ceil(N/2) + zcut);
    figure, imshow(abs(object_xy), [])
    set(gca, 'fontsize', 12, 'xtick', 0:floor(N/4):N, 'ytick',
0:floor(N/4):N, 'ydir', 'normal')
    axis([0 N 0 N]), axis on
    colormap('gray'), colormap(1 - colormap), colorbar
    xlabel('x', 'fontsize',16)
    ylabel('y', 'fontsize',16)
%%%%%%%%%%%%%%%%%%%%%%%%%%%%%%%%%%%%%%%%%%%%%%%%%%%%%%%%%%%%%%%%%%%%%%%%%%%%%%
```

## 10.4 Subroutine: Propagation of wavefront

```
Save it as "Propagator.m" into your "C/Program Files/MATLAB..." folder
%%%%%%%%%%%%%%%%%%%%%%%%%%%%%%%%%%%%%%%%%%%%%%%%%%%%%%%%%%%%%%%%%%%%%%%%%%%%%%
% WAVEFRONT PROPAGATION BY ANGULAR SPECTRUM METHOD
%%%%%%%%%%%%%%%%%%%%%%%%%%%%%%%%%%%%%%%%%%%%%%%%%%%%%%%%%%%%%%%%%%%%%%%%%%%%%%
% Citation for this code/algorithm or any of its parts:
% Tatiana Latychevskaia and Hans-Werner Fink
% "Practical algorithms for simulation and reconstruction of digital in-
line holograms",
% Appl. Optics 54, 2424 - 2434 (2015)
%%%%%%%%%%%%%%%%%%%%%%%%%%%%%%%%%%%%%%%%%%%%%%%%%%%%%%%%%%%%%%%%%%%%%%%%%%%%%%
% The code is written by Tatiana Latychevskaia, 2002
% The version of Matlab for this code is R2010b

function [p] = Propagator(N, lambda, area, z)

p = zeros(N,N);

for ii = 1:N;
    for jj = 1:N
        alpha = lambda*(ii - N/2 -1)/area;
        beta = lambda*(jj - N/2 -1)/area;
        if ((alpha^2 + beta^2) <= 1)
        p(ii,jj) = exp(-2*pi*i*z*sqrt(1 - alpha^2 - beta^2)/lambda);
        end; % if
    end
end;
%%%%%%%%%%%%%%%%%%%%%%%%%%%%%%%%%%%%%%%%%%%%%%%%%%%%%%%%%%%%%%%%%%%%%%%%%%%%%%
```

## 10.5 Subroutine: 2d centered Fourier transform

Save it as "FT2Dc.m" into your "C/Program Files/MATLAB..." folder

```
%%%%%%%%%%%%%%%%%%%%%%%%%%%%%%%%%%%%%%%%%%%%%%%%%%%%%%%%%%%%%%%%%%%%%%%%%%%
% 2d centered Fourier transform
%%%%%%%%%%%%%%%%%%%%%%%%%%%%%%%%%%%%%%%%%%%%%%%%%%%%%%%%%%%%%%%%%%%%%%%%%%%
% Citation for this code and algorithm:
% Tatiana Latychevskaia and Hans-Werner Fink
% "Practical algorithms for simulation and reconstruction of digital in-
line holograms",
% Appl. Optics 54, 2424 - 2434 (2015)
%%%%%%%%%%%%%%%%%%%%%%%%%%%%%%%%%%%%%%%%%%%%%%%%%%%%%%%%%%%%%%%%%%%%%%%%%%%
% The code is written by Tatiana Latychevskaia, 2002
% The version of Matlab for this code is R2010b

function [out] = FT2Dc(in)

[Nx Ny] = size(in);

f1 = zeros(Nx,Ny);

for ii = 1:Nx
    for jj = 1:Ny
         f1(ii,jj) = exp(i*pi*(ii + jj));
    end
end

FT = fft2(f1.*in);

out = f1.*FT;
%%%%%%%%%%%%%%%%%%%%%%%%%%%%%%%%%%%%%%%%%%%%%%%%%%%%%%%%%%%%%%%%%%%%%%%%%%%
```

## 10.6 Subroutine: 2d centered inverse Fourier transform

Save it as “IFT2Dc.m” into your “C/Program Files/MATLAB...” folder

```
%%%%%%%%%%%%%%%%%%%%%%%%%%%%%%%%%%%%%%%%%%%%%%%%%%%%%%%%%%%%%%%%%%%%%%%%%%%
% 2d centered inverse Fourier transform
%%%%%%%%%%%%%%%%%%%%%%%%%%%%%%%%%%%%%%%%%%%%%%%%%%%%%%%%%%%%%%%%%%%%%%%%%%%
% Citation for this code and algorithm:
% Tatiana Latychevskaia and Hans-Werner Fink
% "Practical algorithms for simulation and reconstruction of digital in-
line holograms",
% Appl. Optics 54, 2424 - 2434 (2015)
%%%%%%%%%%%%%%%%%%%%%%%%%%%%%%%%%%%%%%%%%%%%%%%%%%%%%%%%%%%%%%%%%%%%%%%%%%%
% The code is written by Tatiana Latychevskaia, 2002
% The version of Matlab for this code is R2010b

function [out] = IFT2Dc(in)

[Nx Ny] = size(in);

f1 = zeros(Nx,Ny);

for ii = 1:Nx
    for jj = 1:Ny
        f1(ii, jj) = exp(-i*pi*(ii + jj));
    end
end

FT = ifft2(f1.*in);

out = f1.*FT;
%%%%%%%%%%%%%%%%%%%%%%%%%%%%%%%%%%%%%%%%%%%%%%%%%%%%%%%%%%%%%%%%%%%%%%%%%%%
```

## 10.7 Subroutine: 3d centered Fourier transform

```
Save it as "FT3Dc.m" into your "C/Program Files/MATLAB..." folder

%%%%%%%%%%%%%%%%%%%%%%%%%%%%%%%%%%%%%%%%%%%%%%%%%%%%%%%%%%%%%%%%%%%%%%%%%%%%%
% 3d centered Fourier transform
%%%%%%%%%%%%%%%%%%%%%%%%%%%%%%%%%%%%%%%%%%%%%%%%%%%%%%%%%%%%%%%%%%%%%%%%%%%%%
% Citation for this code and algorithm:
% Tatiana Latychevskaia and Hans-Werner Fink
% "Practical algorithms for simulation and reconstruction of digital in-
line holograms",
% Appl. Optics 54, 2424 - 2434 (2015)
%%%%%%%%%%%%%%%%%%%%%%%%%%%%%%%%%%%%%%%%%%%%%%%%%%%%%%%%%%%%%%%%%%%%%%%%%%%%%
% The code is written by Tatiana Latychevskaia, 2002
% The version of Matlab for this code is R2010b

function [out] = FT3Dc(in)

[Nx Ny Nz] = size(in);

f1 = zeros(Nx,Ny,Nz);

for ii = 1:Nx
    for jj = 1:Ny
          for kk = 1:Nz
              f1(ii,jj,kk) = exp(i*pi*(ii + jj + kk));
          end
    end
end

FT = fftn(f1.*in);

out = f1.*FT;
%%%%%%%%%%%%%%%%%%%%%%%%%%%%%%%%%%%%%%%%%%%%%%%%%%%%%%%%%%%%%%%%%%%%%%%%%%%%%
```

## 10.8 Subroutine: 3d centered inverse Fourier transform

```
Save it as "IFT3Dc.m" into your "C/Program Files/MATLAB..." folder

%%%%%%%%%%%%%%%%%%%%%%%%%%%%%%%%%%%%%%%%%%%%%%%%%%%%%%%%%%%%%%%%%%%%%%%%%%%
% 3d centered inverse Fourier transform
%%%%%%%%%%%%%%%%%%%%%%%%%%%%%%%%%%%%%%%%%%%%%%%%%%%%%%%%%%%%%%%%%%%%%%%%%%%
% Citation for this code and algorithm:
% Tatiana Latychevskaia and Hans-Werner Fink
% "Practical algorithms for simulation and reconstruction of digital in-
line holograms",
% Appl. Optics 54, 2424 - 2434 (2015)
%%%%%%%%%%%%%%%%%%%%%%%%%%%%%%%%%%%%%%%%%%%%%%%%%%%%%%%%%%%%%%%%%%%%%%%%%%%
% The code is written by Tatiana Latychevskaia, 2002
% The version of Matlab for this code is R2010b

function [out] = IFT3Dc(in)

[Nx Ny Nz] = size(in);

f1 = zeros(Nx,Ny,Nz);

for ii = 1:Nx
    for jj = 1:Ny
        for kk = 1:Nz
        f1(ii, jj, kk) = exp(-i*pi*(ii + jj + kk));
        end
    end
end

FT = ifftn(f1.*in);

out = f1.*FT;
%%%%%%%%%%%%%%%%%%%%%%%%%%%%%%%%%%%%%%%%%%%%%%%%%%%%%%%%%%%%%%%%%%%%%%%%%%%
```